\documentclass[acmsmall]{acmart}

\usepackage{booktabs} 
\usepackage{hyperref}   
\usepackage{url}  
\usepackage{multirow} 
\usepackage{color}

\usepackage{diagbox}
\usepackage{amsmath}
\usepackage{arydshln} 
\usepackage{color}
\usepackage{algorithmic}
\usepackage{soul}
\usepackage{color, xcolor}  
\usepackage{tabularx}
\usepackage{makecell}

\usepackage[linesnumbered,ruled]{algorithm2e} % For algorithms

\SetAlFnt{\small}
\SetAlCapFnt{\small}
\SetAlCapNameFnt{\small}
\SetAlCapHSkip{0pt}
\IncMargin{-\parindent}

\sethlcolor{yellow}
\soulregister{\cite}7
\soulregister{\citep}7 % 注册\citep命令
\soulregister{\citet}7 % 注册\citet命令
\soulregister{\ref}7 % 注册\ref命令
\soulregister{\pageref}7 % 注册\pageref命令
\soulregister{\underline}7 % 注册\underline命令

\usepackage[switch,pagewise]{lineno}
\acmJournal{TIST}

\setcopyright{acmlicensed}
\acmDOI{TIST}
\begin{document}

% Title portion
\title{Graph Contrastive Learning on Multi-label Classification for Recommendations}

\author{Jiayang Wu}
\affiliation{ %College of Cyber Security, 
	\institution{Jinan University}
	\city{Guangzhou}
	\country{China}
}
\email{csjywu1@gmail.com}

\author{Wensheng Gan}
\authornote{This is the corresponding author.}
\affiliation{
	\institution{Jinan University}
	\city{Guangzhou}
	\country{China}
}
\email{wsgan001@gmail.com}

\author{Huashen Lu}
\affiliation{ %College of Cyber Security, 
	\institution{Jinan University}
	\city{Guangzhou}
	\country{China}
}
\email{huashen.lu@gmail.com}

\author{Philip S. Yu}
\affiliation{
	\institution{University of Illinois Chicago}
	\city{Chicago}
	\country{USA}
}
\email{psyu@uic.edu}

\begin{abstract}
    In business analysis, providing effective recommendations is essential for enhancing company profits. The utilization of graph-based structures, such as bipartite graphs, has gained popularity for their ability to analyze complex data relationships. Link prediction is crucial for recommending specific items to users. Traditional methods in this area often involve identifying patterns in the graph structure or using representational techniques like graph neural networks (GNNs). However, these approaches encounter difficulties as the volume of data increases. To address these challenges, we propose a model called Graph Contrastive Learning for Multi-label Classification (MCGCL). MCGCL leverages contrastive learning to enhance recommendation effectiveness. The model incorporates two training stages: a main task and a subtask. The main task is holistic user-item graph learning to capture user-item relationships. The homogeneous user-user (item-item) subgraph is constructed to capture user-user and item-item relationships in the subtask. We assessed the performance using real-world datasets from Amazon Reviews in multi-label classification tasks.  Comparative experiments with state-of-the-art methods confirm the effectiveness of MCGCL, highlighting its potential for improving recommendation systems. 
\end{abstract}

%
% The code below should be generated by the tool at
% http://dl.acm.org/ccs.cfm
% Please copy and paste the code instead of the example below.
%

%H.2.8 [Database Applications]: Data mining


\maketitle

\renewcommand{\shortauthors}{J. Wu \textit{et al.}}

\section{Introduction}  \label{sec: introduction}

\begin{figure}[ht]
    \centering
    \includegraphics[clip,scale=0.32]{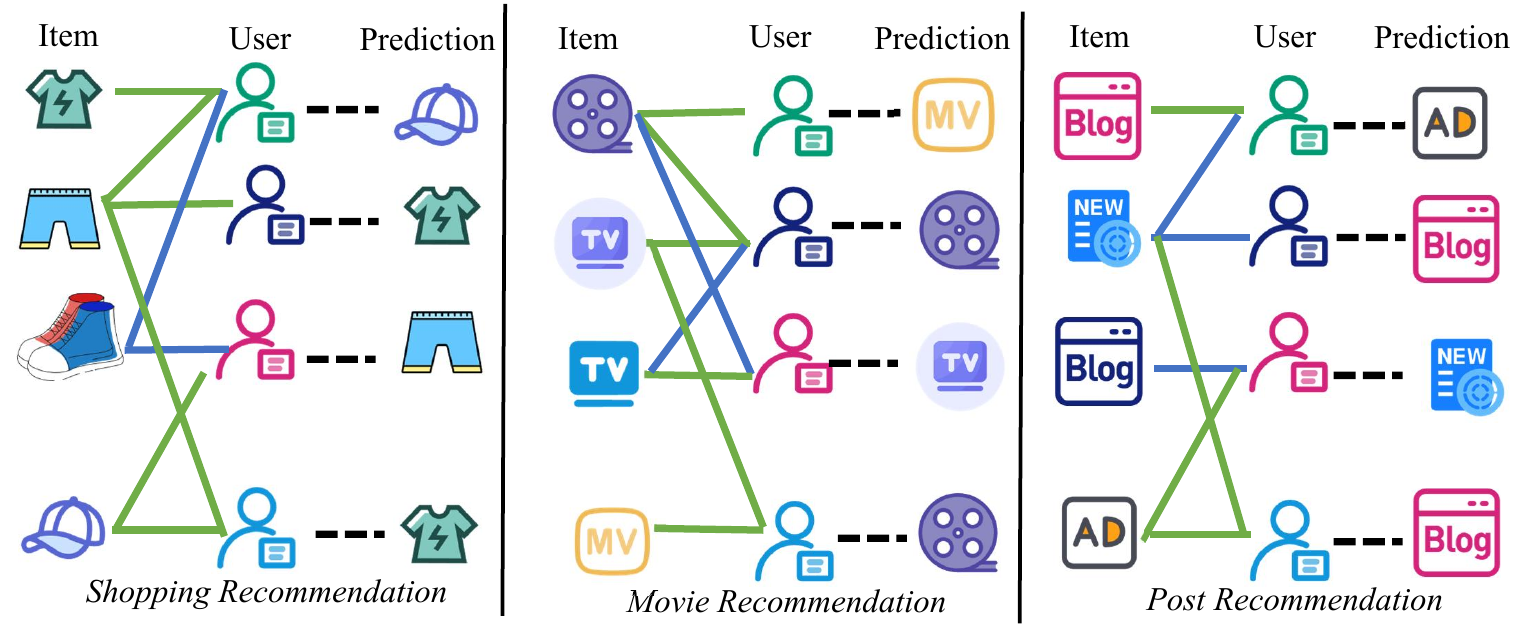}
    \caption{Illustrative scenarios of link prediction.}
    \label{fig:link}
\end{figure}

In the fields of data intelligence \cite{gan2021survey,he2024mining} and personalized recommendations \cite{shani2011evaluating,chen2024data}, transforming scenarios into the bipartite graph format is crucial for understanding user preferences and behaviors. Signed bipartite graphs can map relationships between two groups, such as customers and products, capturing both positive and negative interactions. The link prediction task is one of the most common tasks for these graphs \cite{zhang2020learning}. It can be regarded as the recommendation system to anticipate whether the items should be recommended to corresponding users, as shown in Fig. \ref{fig:link}, where the green line represents the positive rating, whereas the blue line indicates the negative rating.  

Bipartite graphs often include two main methods for learning patterns. One type uses constraint rules to investigate graph structures \cite{iyer2018asap}. The other type employs representation techniques with graph neural networks (GNN) to learn node representations \cite{wu2020comprehensive}. The rule-based method involves using constraint rules to analyze the graph's structure. It can perform well when the structure complexity is simple \cite{iyer2018asap}. However, as the complexity increases, the GNN-based method becomes more advantageous. The advantages of GNNs in processing graphs include utilizing the relationships between nodes and global structural information and adapting to graphs of different scales and dynamic changes. To address the challenge of data sparsity, contrastive learning has gained attention for use in graph recommendation systems. SBGNN is the first algorithm to focus on sub-view contrastive learning in the bipartite graphs  \cite{huang2021signed}. Moreover, SBGCL introduced further improvements by emphasizing the balance between holistic-view and sub-view relationships and leveraging dual perspectives to enrich the representation space \cite{zhang2023contrastive}.

Multi-label data reflects a more accurate reflection of real-world business scenarios than the simpler binary classification approach \cite{lin2023thresholding}. When dealing with multiple labels, establishing sub-view relationships becomes particularly challenging. To address these challenges, our method is inspired by multi-task learning \cite{zhang2021survey} and incorporates homogeneous subgraph learning as a subtask into our framework \cite{yang2023adatask}. Specifically, we obtain the primary representations from the main task, which are then enhanced during the secondary insights in the subtask. 

To address the challenge mentioned above, we propose a model called Graph Contrastive Learning for Multi-label Classification (MCGCL). Our framework decomposes the user-item bipartite graph into one holistic graph and two homogeneous subgraphs. It then performs two tasks: holistic user-item graph learning and homogeneous user-user (item-item) subgraph learning. In the first stage, we focus on learning representations from the holistic graph. In the next stage, the representations obtained from the main task serve as the preliminary knowledge for the subtask. These representations are then utilized to construct the homogeneous subgraph and learn representations in the subgraph. Finally, the third stage involves the integration of representations from both the main task and the subtask. Our work addresses a significant gap in the literature by exploring the application of contrastive learning to multi-label classification in graph data. To our knowledge, no previous research has applied contrastive learning to this problem. Our proposed framework, named MCGCL, employs contrastive learning specifically for multi-label classification in bipartite graphs.  Our contributions to the field are manifold, encompassing four key areas:

\begin{itemize}
    \item  It's the first framework using contrastive learning in the bipartite graph with multi-label classification. We adopt the view of holistic bipartite graph learning and homogeneous subgraph learning and construct graphs with two augmentation methods, including adding edges and removing edges. 
	
    \item We decompose the problem into two tasks and three stages. The main task focuses on obtaining representations from the holistic bipartite graph, while the subtask is designed to learn about the hard sample in the homogeneous subgraph. The third stage involves aggregating the representations from both tasks and merging information from both views.
	
    \item  In the subtask, we utilized the representations generated during the main task as a foundation for constructing the homogeneous subgraph including user-user and item-item graphs. Through this additional view, we successfully acquired representations for hard samples.
	
    \item  We employed the Amazon Reviews datasets to tackle the multi-label classification task and conduct experiments. The results show that MCGCL surpasses the other methods in both multi-label and binary-label classification tasks. Additionally, through an ablation study and analysis of hyperparameters, we further confirmed the robust generalization ability of our model.
\end{itemize}

The remaining parts of this paper are given as follows. Related work is stated and summarized in Section \ref{sec: relatedwork}. The preliminaries and basic knowledge are described in Section \ref{sec: preliminaries}. Our method MCGCL is detailed in Section \ref{sec: algorithm}. Furthermore, we conducted some experiments, and the experimental results are shown in Section \ref{sec: experiments}. Finally, we present the conclusion in Section \ref{sec: conclusion}.

\section{Related Work} \label{sec: relatedwork}
\subsection{Graph representation learning}

Graph representation learning is a powerful technique for understanding the relationships between entities in graph-based data \cite{wu2024mdgrl}. By turning entities and relationships into vectors, we can represent them as numerical values in a smaller-dimensional space. This allows us to capture their meaning and attributes. These representations enable us to measure similarities between entities, uncovering hidden connections. In the field of recommendation systems, graph representation learning (GRL) has become an essential method for capturing complex relationships between users and items. GERL utilizes a transformer architecture to construct semantic representations of news and enhances these representations by incorporating information from neighboring news through a graph attention network \cite{ge2020graph}. Wang \textit{et al.} \cite{wang2020m2grl} proposed M2GRL, which builds a graph for each single-view data and learns multiple independent representations from these graphs, improving the performance of large-scale recommendation systems. Additionally, Zhang \textit{et al.} \cite{zhang2021blog} proposed BLoG, which combines local and global regularization and optimizes the graph encoders, thereby enhancing the extraction of graph structural features and recommendation performance. Moreover, Wang \textit{et al.} \cite{wang2020group} presented the GLS-GRL for sequential group recommendation, which constructs group-aware long-term and short-term graphs and learns and integrates user representations through graph representation learning.

\subsection{Graph contrastive learning}

Nowadays, graph data tend to be sparse and often contain noise. In response to these challenges, various methods have been developed, including the application of contrastive learning to alleviate them \cite{chen2020simple, zhu2021graph}. Graph augmentation is the core of contrastive learning \cite{zhao2021data}. This involves variations in graph structures or applying random perturbations. Graph contrastive learning (GCL) enhances recommendation systems by integrating structural and semantic graph information through self-supervised learning. HGCL \cite{chen2023heterogeneous} leverages heterogeneous relational data using contrastive learning and personalized data augmentation to improve user and item representations. ADAGCL \cite{jiang2023adaptive} employs contrastive self-supervised learning and graph neural networks to model complex user-item relationships and capture temporal dependencies. SimGCL \cite{yu2022graph} simplifies the contrastive learning process by adding uniform noise to the representation space instead of complex graph augmentations. Lastly, KGCL \cite{yang2022knowledge} utilizes knowledge graph information and contrastive learning to enhance user and item representations, handling data noise and sparsity.

\subsection{Usage of sub-view relationship}

\begin{figure}[ht]
    \centering
    \includegraphics[clip,scale=0.32]{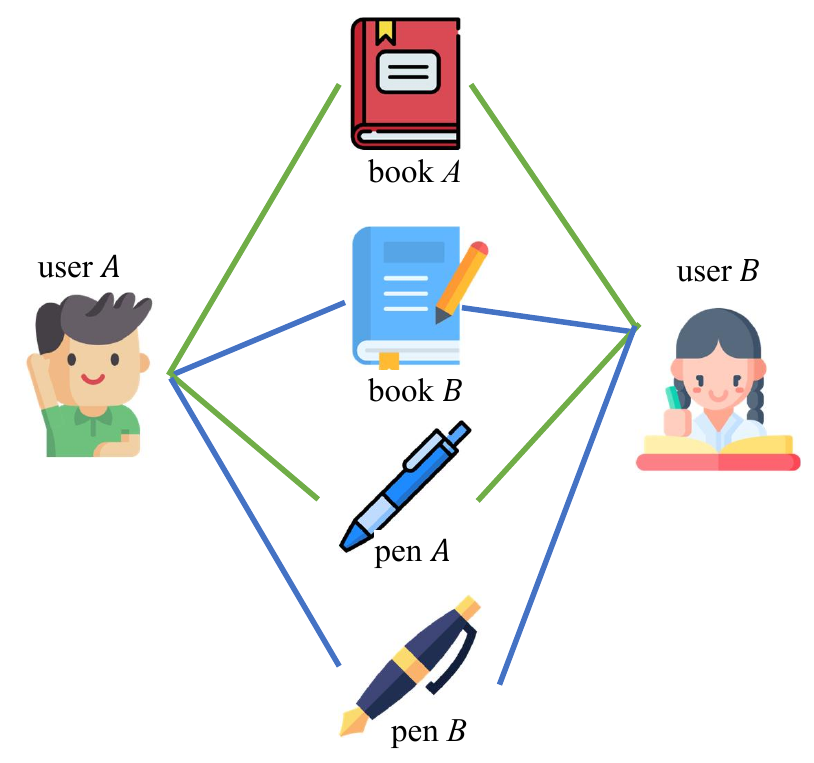}
    \caption{Example of link prediction based on user ratings.}
    \label{fig:sub}
\end{figure}

Focusing only on holistic graph learning may restrict the depth of analysis. Thus, establishing sub-view relationships in each set is also essential. For example, as shown in Fig. \ref{fig:sub}, given a bipartite graph,  if user $A$ and user $B$ both give a positive rating to pen $A$  and book $A$ and a negative rating to pen $B$ and book $B$, it suggests that user $A$ and user $B$ have similar preference. As a result, the proximity between user $A$ and user $B$ is regarded as close, indicating a positive relationship.  Some methods have applied the principles of balance theory to define relationships by using the structural patterns of graphs \cite{wang2017signed,sun2023efficient,guo2023social}. The triangle and butterfly theorems serve as effective strategies for forming relationships and resolving inconsistencies in the homogeneous subgraph. However, these approaches may not perform well with sparse data, leading to reduced accuracy. To address these challenges, our strategy involves utilizing knowledge of holistic-view relationships to help homogeneous subgraph learning, adopting a multi-task paradigm. The development of the homogeneous subgraph is approached as a sub-task.

\subsection{Multi-label classification}

\begin{figure}[ht]
    \centering
    \includegraphics[clip,scale=0.2]{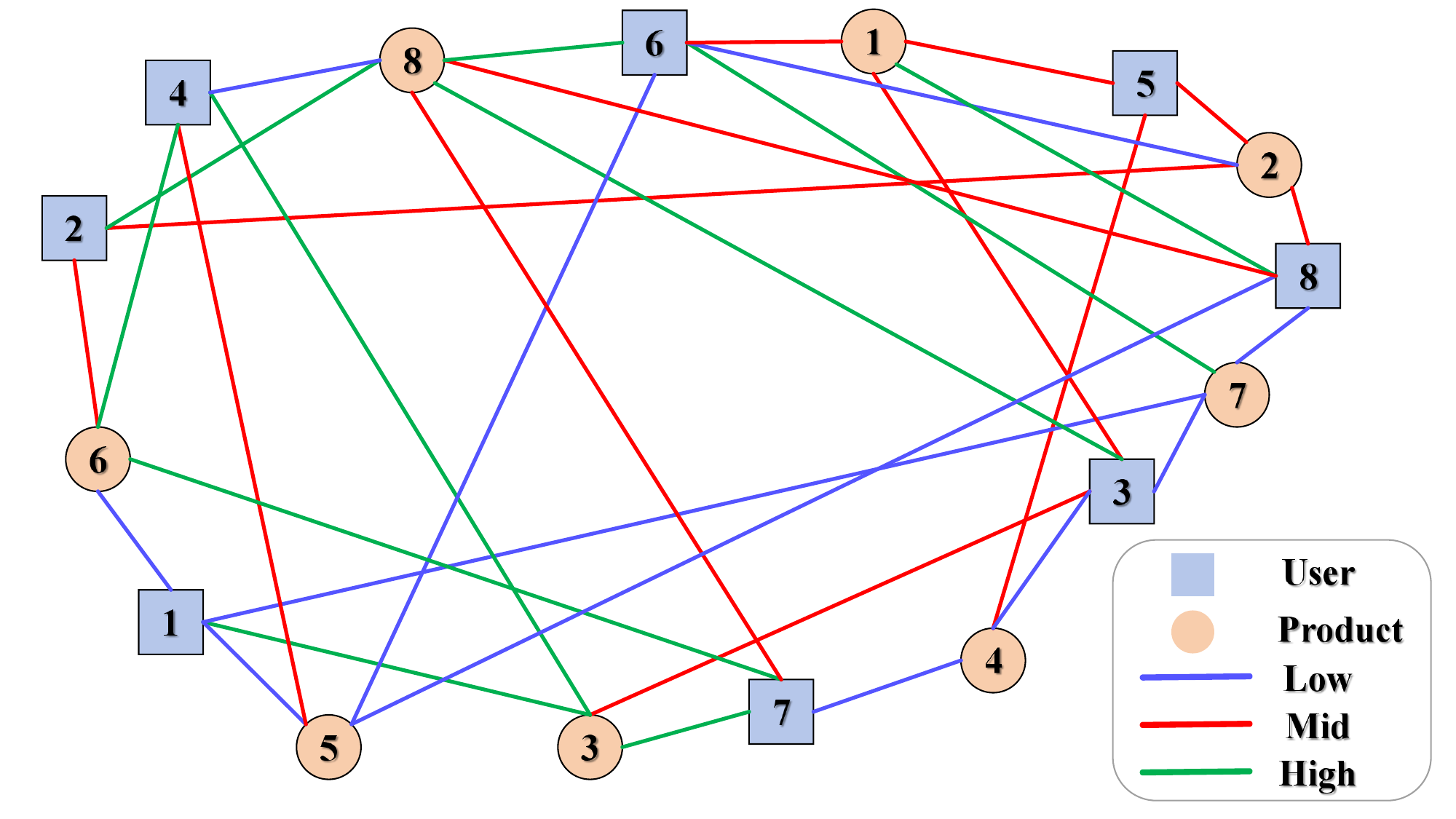}
    \caption{Illustrative diagram of a multi-label bipartite graph.}
    \label{fig:network}
\end{figure}

In real-world applications, data often comes with diverse labels, as shown in Fig. \ref{fig:network}. End-to-end learning is the method where the model learns from input data and outputs predictions without the need for explicit feature extraction. By integrating all processing steps into a single model, it can automatically capture complex patterns in the data. NeuLP \cite{zhong2020neulp} addresses the limitations of GNNs by integrating the linear and nonlinear properties of GNNs, effectively utilizing user attributes and interactions. The SEAL algorithm \cite{cai2021line} uses line graph transformation. It avoids information loss in pooling layers and reduces computational complexity. Additionally, it enhances prediction accuracy by converting subgraphs to line graphs. The M-GNN \cite{wang2019robust} introduces a multi-level GNN. Its encoder embeds entities by aggregating information from neighboring nodes, and the decoder uses the learned embeddings to compute edge probabilities.

Contrastive learning is an unsupervised or self-supervised learning method that learns useful representations of data. It achieves this by maximizing the consistency between similar samples and minimizing the consistency between different samples. For example,  the LRDG \cite{zhang2024multi} improves multi-label classification by discovering and generating latent relationships. It enhances by generating new data samples to help the model better understand label relationships. This approach is particularly suitable for handling sparse label data. C-GMVAE \cite{bai2022gaussian} combines contrastive learning with Gaussian mixture variational autoencoders (GMVAE) to enhance multi-label classification performance. It achieves this by better capturing data distribution through the use of a Gaussian mixture model and enhancing the correlation between labels and features. This approach demonstrates significant advantages on high-dimensional datasets. The MulSupCon algorithm \cite{zhang2024multi}, combines supervised contrastive learning to handle multi-label classification. It defines positive and negative samples more accurately through a new loss function. It considers label overlap and uses the contrastive loss function that measures the similarity between samples.

In multi-label classification tasks, contrastive learning has several notable advantages over end-to-end learning methods: first, contrastive learning can better capture complex relationships between labels by bringing similar samples closer together and pushing different samples apart. This can significantly enhance the ability to capture complex relationships. Second, label imbalance is a common issue in multi-label classification, and contrastive learning can alleviate this problem by optimizing the representation space, ensuring reasonable representation for each label.

\section{Preliminaries}  \label{sec: preliminaries}
In this section,  we first provide the formal problem statement of the multi-label classification of link prediction in the bipartite graph. Then, the definitions and fundamental concepts used in this paper are introduced.

\begin{definition}[Problem definition]
    \rm  The bipartite graph data for recommendation can be represented as $G$ = $(V, E)$, where $V$ = $\{U \cup I \}$. $U$ denotes the set of users \{$u_{1}$, $u_{2}$, ..., $u_{{_{|U|}}}$\} and $I$ represents the set of items \{$i_{1}$, $i_{2}$, ..., $i_{_{|I|}}$\}. The edges, denoted as $\varepsilon \in E$, establish connections between users and items. These edges are categorized into three rating labels: high ($\varepsilon ^h$), medium ($\varepsilon ^m$), and low ($\varepsilon ^l$), reflecting the degree of user satisfaction towards the items, including high, medium and low. The core objective of our study is to predict the label of $\varepsilon$ that exists between a user $u$ and an item $i$, formally described as $(u, i) \in \,\,\varepsilon ^{?}$. 
    
    We extract the edges and partition the graph $G$ into subsets: $G_{\textit{train}}$ for training and $G_{\textit{test}}$ for testing.  We design a framework that learns the patterns from the holistic graph and homogeneous graph from $G_{\textit{train}}$. Moreover, we incorporate a validation set to combine the representations between two views. Finally, we will assess the effectiveness of our model in $G_{\textit{test}}$.
\end{definition}

\begin{definition}[Graph encoder layer]
    \rm Our framework employs graph convolutional networks (GCNs) for graph encoding \cite{wu2019simplifying}. Each node is a vector $d$-dimensional vector, where both users and items are denoted by $h_{_{\mathcal{U}(\mathcal{I})}\,\,} \in \mathbb{R} ^d$. The principle of GCNs is message propagation, which allows nodes to aggregate features from their neighboring nodes. This process involves iterative node aggregation across multiple layers. After this multi-layered process, the representation of a user node at the $k$-th layer as $h_{u}^{\left( k \right)}$ can be formulated as: 
    \begin{equation} \label{eq1}
        a_{v}^{\left( k \right)}=\textit{AGGREGATE}^{\left( k \right)}\left(\left( h_{v}^{\left( k-1 \right)} \right) , \forall v \in N \left( u \right) \right)
    \end{equation}

    \begin{equation} \label{eq2}
        h_{u}^{\left( k \right)} = \textit{COMBINE}^{\left( k \right)}\left( h_{u}^{\left( k-1 \right)},a_{v}^{\left( k \right)} \right) 
    \end{equation}
    The \textit{AGGREGATE} function gathers features from adjacent nodes, while the \textit{COMBINE} function integrates the information from the $(k-1)$-th layer to construct $a_{v}^{\left( k \right)}$. From the perspective of layers, the operation can be described as follows:
     \begin{equation} \label{eq3}
        H^k = \textit{softmax} \left( \tilde{A}H^{k-1}W^{k-1} \right) 
     \end{equation}
\end{definition}

\begin{figure}[ht]
    \centering
    \includegraphics[clip,scale=0.45]{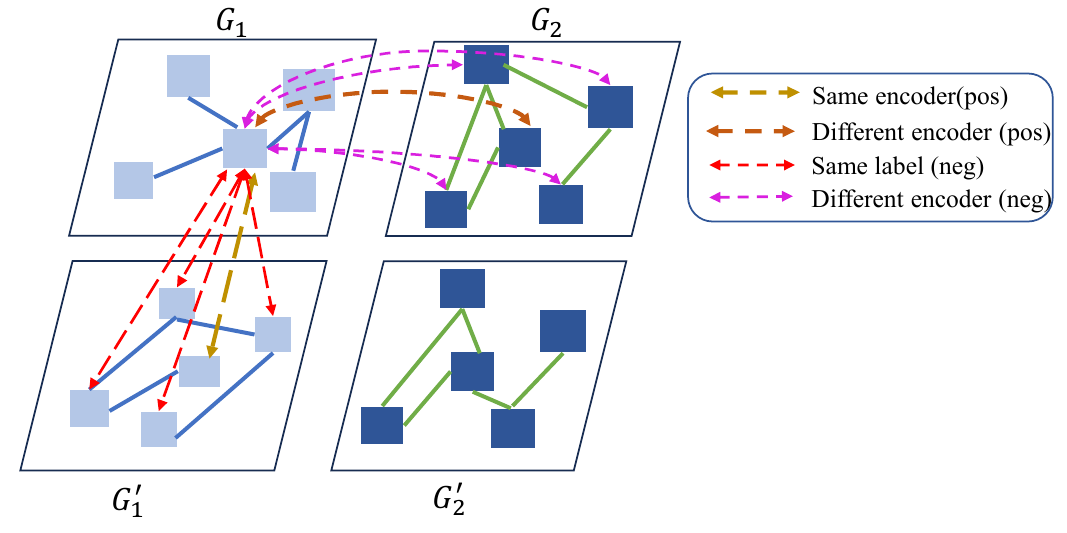}
    \caption{Comparative analysis using the same encoder for $G_{1}, G^{'}_{1}$, and using different encoders for $G_{1}, G_{2}$.}
    \label{fig:loss}
\end{figure}

\begin{definition}[Contrastive loss function]
     \rm The comparison of graph representations depends on whether the sets of graphs are processed by the same encoder. This is illustrated in Figure \ref{fig:loss}.  When using the same encoder, it focuses on comparing the augmented graph with the same labels, such as $G_{1}$ with $G_{1}^{'}$ and $G_{2}$ with $G_{2}^{'}$. For instance, the contrastive loss between $G_{1}$ and its augmented graph $G_{1}^{'}$, both carrying high labels, can be calculated as follows \cite{zhang2023contrastive}:
     \begin{small}
    \begin{equation} \label{500}
        L^{\textit{same}}= -\frac{1}{\mathcal{|I|}}\sum_{i=1}^\mathcal{_{\mathcal{|I|}}}{\log \frac{\exp \left( sim\left( H_{_{i},_{G_1}}^{h},H_{_{i,G_{1}^{'}}}^{h} \right) \right)}{\sum_{j=1,j\ne i}^\mathcal{_{\mathcal{|I|}}}{\exp \left( sim\left( H_{_{i,G_1}}^{h},H_{_{j,G_{1}^{'}}}^{h} \right) \right)}}}-\frac{1}{\mathcal{|U|}}\sum_{u=1}^{_{\mathcal{|U|}}}{\log \frac{\exp \left( sim\left( H_{_{u,G_1}}^{h},H_{_{u,G_{1}^{'}}}^{h} \right) \right)}{\sum_{u=1,u\ne v}^{_{\mathcal{|U|}}}{\exp \left( sim\left( H_{_{u,G_1}}^{h},H_{_{v,G_{1}^{'}}}^{h} \right) \right)}}} 
    \end{equation}
    \end{small}
     where $\mathcal{|I|}$ is the number of items and $\mathcal{|U|}$ is the number of users, $H$ represents the nodes' embeddings, $Z_{i,G_{1}}$ represents the $i$-th node in the contrastive graph $G_{1}$, and $Z_{j,G_{1}^{'}}$ represents the rest of the nodes in the augmented graph $G_{1}^{'}$. The variables $i$ and $j$ are different, meaning that $i\neq j$. This $\textit{sim}(\cdot,\cdot)$ represents the similarity function between the two representations by using cosine similarity.
     
    Moreover, when comparing graphs with different labels, this can also be interpreted as contrasting the representations produced by different encoders. Taking the contrastive loss between $G_{1}$ and $G_{2}$ as an illustration, its calculation is similar to Formula \ref{500}. The following show the calcautions between high labels and low labels:
    \begin{small}
        \begin{equation} \label{501}
        L^{\textit{differ}}= -\frac{1}{\mathcal{|I|}}\sum_{i=1}^{_{\mathcal{|I|}}}{\log \frac{\exp \left( sim\left( H_{_{i,G_1}}^{h},H_{_{i,G_2}}^{l} \right) \right)}{\sum_{j=1,j\ne i}^{_{\mathcal{|I|}}}{\exp \left( sim\left( H_{_{i,G_1}}^{h},H_{_{j,G_2}}^{l} \right) \right)}}}-\frac{1}{\mathcal{|U|}}\sum_{u=1}^{_{\mathcal{|U|}}}{\log \frac{\exp \left( sim\left( H_{_{u,G_1}}^{h},H_{_{u,G_2}}^{l} \right) \right)}{\sum_{u=1,u\ne v}^{_{\mathcal{|U|}}}{\exp \left( sim\left( H_{_{u,G_1}}^{h},H_{_{v,G_{2}}}^{l} \right) \right)}}}
        \end{equation}
    \end{small}
\end{definition}

\begin{definition}[Homogeneous graph and holistic graph]
    \rm  A homogeneous graph \cite{wan2022inductive}  is characterized by nodes and edges of a single type. This means all nodes represent the same kind of entity, and all edges denote the same kind of relationship. For example, in a graph where all nodes represent users and all edges indicate interactions between users, the graph is considered homogeneous. In contrast, a holistic graph \cite{rassil2022holistic} is a more complex structure that encompasses multiple types of nodes and edges. This allows for a richer representation of diverse entities and their relationships. For instance, a holistic graph might include both user and product nodes. The edges can represent various interactions, such as user-product interactions.
\end{definition}

\section{Algorithm} \label{sec: algorithm}
\subsection{The overall architecture of  MCGCL}

\begin{figure}[ht]
    \centering
    \includegraphics[clip,scale=0.45]{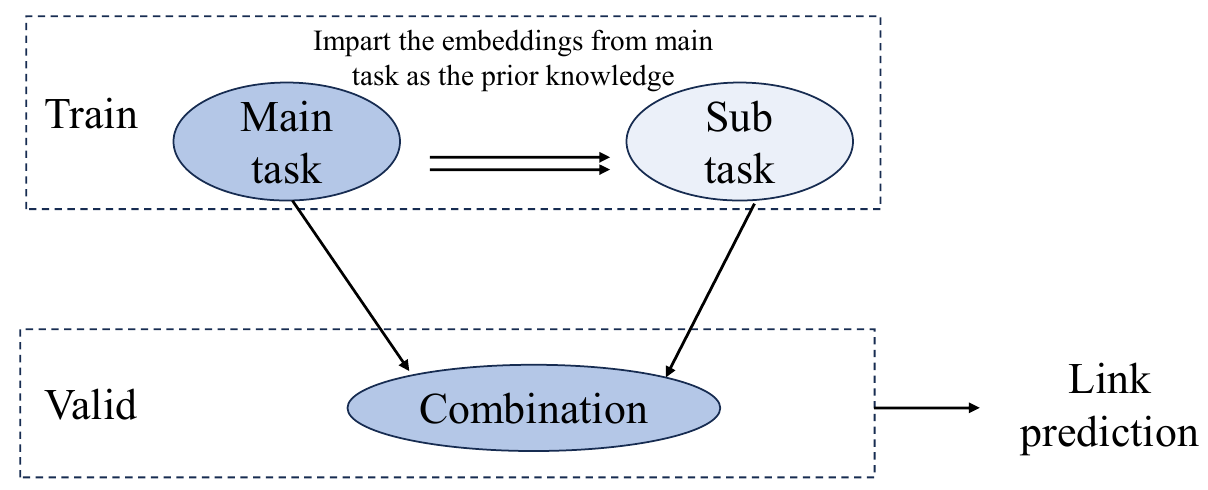}
    \caption{Overview of the training process.}
    \label{fig:abstract}
\end{figure}

The framework begins with a main task. It focuses on learning node representations from the whole bipartite graph, specifically within the user-item context.
Then in the subtask, these representations are used to establish relationships in the homogeneous user-user (item-item) subgraph. This phase is dedicated to focusing on learning representations for the hard samples. It is important to highlight that these two tasks are interconnected. The main task lays the foundation by initializing the representations, while the subtask extracts these representations for hard samples. To combine the two tasks, we employ attention aggregation. The framework can also be reviewed in Fig. \ref{fig:abstract}.

\subsection{Holistic-view for user-item graph}

\begin{figure*}[ht]
    \centering
    \includegraphics[clip,scale=0.52]{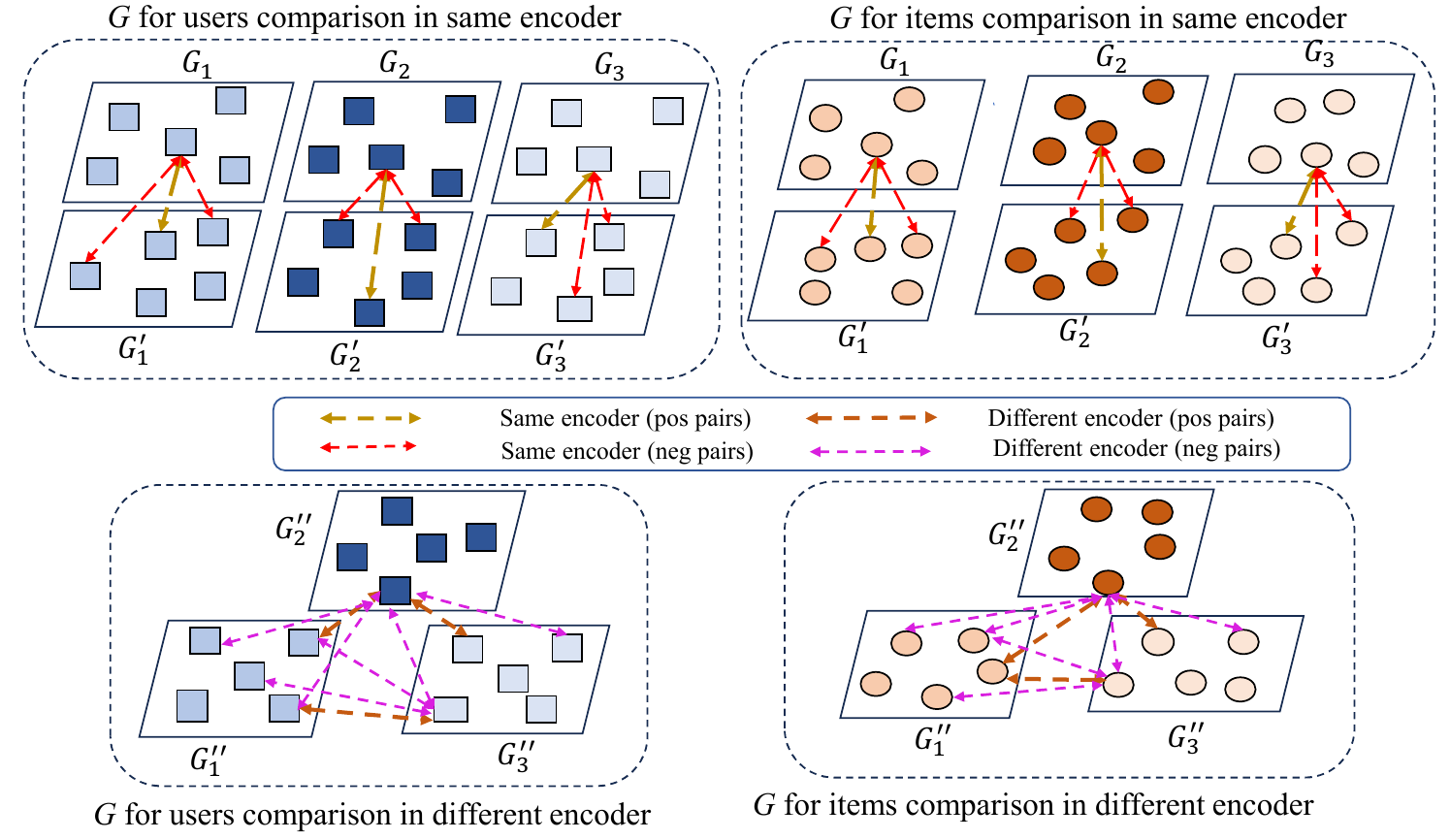}
    \caption{The upper section illustrates contrastive learning in the same encoders for users and items, while the lower section details contrastive learning across different encoders.}
    \label{fig:loss1}
\end{figure*}

\begin{figure*}[ht]
    \centering
    \includegraphics[clip,scale=0.47]{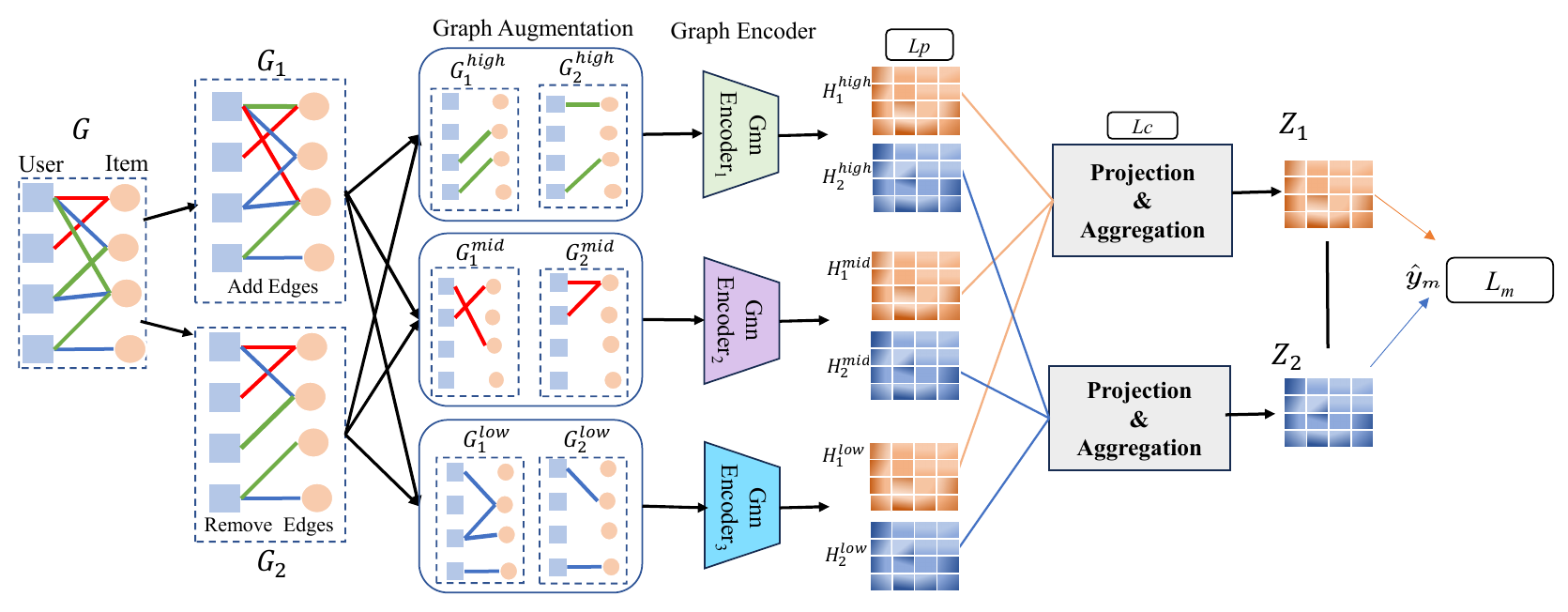}
    \caption{The main task of MCGCL.}
    \label{fig:MCGCL1}
\end{figure*}

The main task of MCGCL is illustrated in Fig. \ref{fig:MCGCL1}. The input graph is denoted as $G$. To enrich the learning process, two augmented graphs, $G_{1}$ and $G_{2}$, are created by applying perturbations. Specifically, $G_{1}$ results from edge removals, and $G_{2}$ from random edge additions, symbolized by $T\in {t,t'}$. These augmented graphs are then classified into three categories based on their labels—high, medium, and low, denoted as \textit{type} = ${h,m,l}$. For each category, three separate $GCN$ encoders vectorize the nodes, sharing parameters in the same label group. For example, the first encoder handles two high-rated augmented graphs, producing different representations for items $<H_{i,t}^{h}$, $H_{i,t'}^{h}>$ and users $<H_{u,t}^{h}$, $H_{u,t'}^{h}>$. Similarly, representations for the medium and low ratings are generated, including items $<$$H_{i,t}^{m}$, $H_{i,t'}^{m}$$>$, $<$$H_{i,t}^{l}$, $H_{i,t'}^{l}$$>$ and users $<$$H_{u,t}^{m}$, $H_{u,t'}^{m}$$>$, $<$$H_{u,t}^{l}$, $H_{u,t'}^{l}$$>$ This results in two sets of matrices per encoder. Comparing representations from the same encoder for two augmented graphs enables calculating the loss $L_{_{M_{p}}}$, from Formulation \ref{500} and deriving the following loss:
\begin{equation} \label{eq11}
L_{_{M_{p}}}=L^{h}+L^{m}+L^{l}
\end{equation}
where also as outlined in the upper section of Fig. \ref{fig:loss1}.
For each label type, with the original and its augmented graph denoted as $t$ and $t'$ respectively and \textit{type} = $\{h,m,l\}$, the loss is computed as:

\begin{small}
    \begin{equation} \label{eq12}
    L^{\textit{type}}=-\frac{1}{|\mathcal{I}|}\sum_{i=1}^{_{|\mathcal{I|}}}{\log \frac{\exp \left( \textit{sim}\left( H_{i,t}^{\textit{type}},H_{i,t'}^{\textit{type}} \right) \right)}{\sum_{j=1,j\ne i}^{_{|\mathcal{I|}}}{\exp \left( \textit{sim}\left( H_{i,t}^{\textit{type}},H_{j,t'}^{\textit{type}} \right) \right)}}}-\frac{1}{|\mathcal{U|}}\sum_{u=1}^{_{|\mathcal{U|}}}{\log \frac{\exp \left( \textit{sim}\left( H_{u,t}^{\textit{type}},H_{u,t'}^{\textit{type}} \right) \right)}{\sum_{v=1,v\ne u}^{_{|\mathcal{U|}}}{\exp \left( \textit{sim}\left( H_{u,t}^{\textit{type}},H_{v,t'}^{\textit{type}} \right) \right)}}}
\end{equation}
\end{small}
 
Given that each encoder generates two matrices, we need to combine the representations from them. For each encoder, the representations are merged using attention \cite{vaswani2017attention}, as follows:
\begin{equation} \label{eq200}
H_{\mathcal{_I}_{,t(t')}}^{\textit{type}} = \frac{1}{|\mathcal{I}|} \sum_{i \in \mathcal{I}} \left( a_{t(t')}^{h} \cdot \tanh(WH_{i,t(t')}^{\textit{type}} + b) \right)
\end{equation}

\begin{equation} \label{eq201}
\beta_{\mathcal{_I}_{,t(t')}} = \frac{\exp \left( H_{\mathcal{_I}_{,t(t')}}^{\textit{type}} \right)}{\exp \left( H_{\mathcal{_I}_{,t'}}^{\textit{type}} \right) + \exp \left( H_{\mathcal{_I}_{,t'}}^{\textit{type}} \right)}
\end{equation}

\begin{equation} \label{eq202}
H_{\mathcal{_I}}^{\textit{type}} = \beta_{\mathcal{_I}_{,t}} H_{\mathcal{_I}_{,t}}^{\textit{type}} + \beta_{\mathcal{_I}_{,t'}} H_{\mathcal{_I}_{,t'}}^{\textit{type}}
\end{equation}
Here, $|\mathcal{I}|$ represents the number of items, $W$ and $b$ are learnable matrices of dimensions $|d| \times |d|$ and $|d| \times 1$, and $a^{h}$ are weights for combining levels of augmentation with dimension $|d| \times 1$. This can help the combination of perspectives from two augmentations.

Following this, for each label category, we obtain sets of user and item representations,  $<$$H_{\mathcal{_U}}^{h}$, $H_{\mathcal{_U}}^{m}$, $H_{\mathcal{_U}}^{l}$$>$ and $<$$H_{\mathcal{_I}}^{h}$, $H_{\mathcal{_I}}^{m}$, $H_{\mathcal{_I}}^{l}$$>$. We then pass through two projection layers to map the representations into a targeted space. Consequently, we acquire sets of projected representations for users and items across different ratings, $<$$Z_{\mathcal{_U}}^{h}$, $Z_{\mathcal{_U}}^{m}$, $Z_{\mathcal{_U}}^{l}$$>$ and $<$$Z_{\mathcal{_I}}^{h}$, $Z_{\mathcal{_I}}^{m}$, $Z_{\mathcal{_I}}^{l}$$>$. As illustrated in the lower part of Fig. \ref{fig:loss1}, we compute the loss $L_{_{M_{c}}}$ for pairs of graphs with different types as follows:

\begin{footnotesize}
\begin{equation} \label{eq203}
L^{\textit{type},\textit{type}^{'}}=-\frac{1}{|\mathcal{I}|}\sum_{i=1}^{|\mathcal{I}|}{\log \frac{\exp \left( sim\left( Z_{i}^{\textit{type}},Z_{i}^{\textit{type}^{'}} \right) \right)}{\sum_{j=1,j\ne i}^{|\mathcal{I}|}{\exp \left( sim\left( Z_{i}^{\textit{type}},Z_{j}^{\textit{type}^{'}} \right) \right)}}}-\frac{1}{|\mathcal{U}|}\sum_{u=1}^{|\mathcal{U}|}{\log \frac{\exp \left( sim\left( Z_{u}^{\textit{type}},Z_{u}^{\textit{type}^{'}} \right) \right)}{\sum_{v=1,v\ne u}^{|\mathcal{U}|}{\exp \left( sim\left( Z_{u}^{\textit{type}},Z_{v}^{\textit{type}^{'}} \right) \right)}}}\,
\end{equation}
\end{footnotesize}
where $\textit{type} \ne \textit{type}^{'}$.
Summing up the graph comparison across all pairs, we obtain:
\begin{equation} \label{eq204}
L_{_{M_{c}}}=\sum{_{\textit{type}}}\sum{_{\textit{type}^{'}\ne \textit{type}}}L^{\textit{type},\textit{type}^{'}}
\end{equation}

Furthermore, we aggregate the user and item representations into $Z_{\mathcal{_U}}^{M}$ and $Z_{\mathcal{_I}}^{M}$. We employ attention aggregation to learn the weighting value. 
The aggregation process, considering three sets of attention weights $\alpha^{(h)}$, $\alpha^{(m)}$, and $\alpha^{(l)}$ corresponding to query matrices $Q_{\mathcal{_{U(I)}}}^{(h)}$, $Q_{\mathcal{_{U(I)}}}^{(m)}$, and $Q_{\mathcal{_{U(I)}}}^{(l)}$ with their respective key matrices $K_{\mathcal{_{U(I)}}}^{(h)}$, $K_{\mathcal{_{U(I)}}}^{(m)}$, and $K_{\mathcal{_{U(I)}}}^{(l)}$, the aggregation process can be represented as follows:
\begin{equation} \label{eq205}
Z_{\mathcal{_{U(I)}}}^{M}=\alpha ^{(h)}\cdot H_{\mathcal{_{U(I)}}}^{(h)}+\alpha ^{(m)}\cdot H_{\mathcal{_{U(I)}}}^{(m)}+\alpha ^{(l)}\cdot H_{\mathcal{_{U(I)}}}^{(l)},
\end{equation}
where $M$ indicates the main task and $\alpha^{(h)}$, $\alpha^{(m)}$, and $\alpha^{(l)}$ are the attention weights derived  from the \textit{softmax} of the dot products of the query and key matrices for each rating \cite{vaswani2017attention}:
\begin{equation} \label{eq206}
\alpha_{\mathcal{_{U(I)}}}^{(h),(m),(l)}=
\textit{softmax} (Q_{\mathcal{_{U(I)}}}^{(h),(m),(l)}\cdot(K_{\mathcal{_{U(I)}}}^{(h),(m),(l)})^T)
\end{equation}
where the query $ Q $ and key $ K $ matrices for each are given by:
\begin{equation} \label{eq207}
Q_{\mathcal{U(I)}}^{(h)} = W_Q^{(h)} H_{\mathcal{U(I)}}^{(h)}, \quad K_{\mathcal{U(I)}}^{(h)} = W_K^{(h)} H_{\mathcal{U(I)}}^{(h)},
\end{equation}
\begin{equation} \label{eq208}
Q_{\mathcal{U(I)}}^{(m)} = W_Q^{(m)} H_{\mathcal{U(I)}}^{(m)}, \quad K_{\mathcal{U(I)}}^{(m)} = W_K^{(m)} H_{\mathcal{U(I)}}^{(m)},
\end{equation}
\begin{equation} \label{eq209}
Q_{\mathcal{U(I)}}^{(l)} = W_Q^{(l)} H_{\mathcal{U(I)}}^{(l)}, \quad K_{\mathcal{U(I)}}^{(l)} = W_K^{(l)} H_{\mathcal{U(I)}}^{(l)}.
\end{equation}

In this context, $ \alpha^{(h)} $, $ \alpha^{(m)} $, and $ \alpha^{(l)} $ serve as attention weights to linearly combine the three input matrices, producing the aggregated output. Utilizing these representations, we can address the main task of predicting the edge labels as follows:
\begin{equation} \label{eq207}
\hat{y}_{_{M_{n,h}}}, \hat{y}_{_{M_{n,m}}}, \hat{y}_{_{M_{n,l}}} = \textit{softmax}\left( \textit{MLP}\left( Z^{M}_{\mathcal{_I}_{,i}} \parallel Z^{M}_{\mathcal{_U}_{,u}} \right) \right),
\end{equation}
Here, we predict the labels for the $ n $-th links in the graph $ G $, where $ Z_{\mathcal{_U}_{,u}} $ and $ Z_{\mathcal{_I}_{,i}} $ denote the representations for the $ u $-th user and the $ i $-th item, drawn from their respective sets $ |\mathcal{U}| $ and $ |\mathcal{I}| $. The $ \textit{MLP}(\cdot) $ constructs a multi-layer perceptron, and the $ \textit{softmax}(\cdot) $ converts the numerical values into a probability distribution. Moreover, the loss function is formulated as:
\begin{equation} \label{eq208}
L_{M} = \sum_{n=1}^{N} y_{_{M_{n}}} \log \left( \hat{y}_{_{M_{n}}} \right) + \eta \left( \sum_{u}^{|\mathcal{U}|} \left\| Z_{\mathcal{U},u}^{M} - Z_{\mathcal{U}}^{M,\textit{read}} \right\|_2^2 + \sum_{i}^{|\mathcal{I}|} \left\| Z_{\mathcal{I},i}^{M} - Z_{\mathcal{I}}^{M,\textit{read}} \right\|_2^2 \right),
\end{equation}
where $ N $ represents the total number of links in $ G $, and $ y_{M_{n}} $ is the true label, which is 1 if the edge exists, and 0 otherwise. The \textit{Readout} operation (specifically mean pooling) is applied on the representations to obtain $ Z^{M,\textit{read}}_{\mathcal{I}} $ and $ Z^{M,\textit{read}}_{\mathcal{U}} $. The loss includes regularization terms to enforce constraints between the representations: $ Z^{M,\textit{read}}_{\mathcal{I}} $ and $ Z^{M}_{\mathcal{I}} $, and between $ Z^{M,\textit{read}}_{\mathcal{U}} $ and $ Z^{M}_{\mathcal{U}} $. These constraints are applied using the squared Euclidean norm (L2 norm squared), denoted as $ \left\| \cdot \right\|_2^2 $. The training steps for this task are outlined in Algorithm \ref{alg:main}.

 \begin{algorithm}
\small
\caption{procedure for the main task}
\label{alg:main}
\begin{algorithmic}[1]
\renewcommand{\algorithmicrequire}{\textbf{Input:}}
\renewcommand{\algorithmicensure}{\textbf{Output:}}
\REQUIRE{bipartite graph $G$; Number of epochs $T$}
\ENSURE{node representations $Z^{M}_{\mathcal{_U}}, Z^{M}_{\mathcal{_I}}$}
\
\STATE{initialize original node representations for high, medium, and low ratings: $H_{\mathcal{_U}}^{h}, H_{\mathcal{_I}}^{h}$, $H_{\mathcal{_U}}^{m}, H_{\mathcal{_I}}^{m}$, $H_{\mathcal{_U}}^{l}, H_{\mathcal{_I}}^{l}$}
\STATE{Set up the $GCN$ encoders with initial parameters for the main task}
\STATE{
generate augmented graphs $G_t$, $G_{t'}$ from $G$
}
\STATE{
segment $G_t$, $G_{t'}$ into $<$$G_{t}^{h}$, $G_{t'}^{h}$$>$, $<$$G_{t}^{m}$, $G_{t'}^{m}$$>$, $<$$G_{t}^{l}$, $G_{t'}^{l}$$>$ based on labels
}
\FOR{\textit{epoch} = 1 to $T$}
\STATE{
obtain representations by training three $GCN$ encoders
}
\STATE{
compute the loss $L_{_{M_{p}}}$ for augmented graphs via Formula \ref{eq11} and \ref{eq12}
}
\STATE{
aggregate representaions using Formula \ref{eq200}, \ref{eq201}, and \ref{eq202}
}
\STATE{
determine the contrastive loss $L_{_{M_{c}}}$ for graphs using Formula \ref{eq203} and \ref{eq204}
}
\STATE{
aggregate the representations into $Z^{{M}}_{\mathcal{_U}}$, $Z^{{M}}_{\mathcal{_I}}$ as specified in Formula \ref{eq205} and \ref{eq206}
}
\STATE{
compute the link prediction loss $L_{_{M}}$ via Formula \ref{eq207} and \ref{eq208}
}
\STATE{perform backpropagation and update the model parameters}
\ENDFOR
\RETURN {$Z^{{M}}_{\mathcal{_{U}}}$, $Z^{{M}}_{\mathcal{_{I}}}$}
\end{algorithmic}
\end{algorithm}

\subsection{Sub-view for user-user \& item-item graph}

\begin{figure*}[ht]
    \centering
    \includegraphics[clip,scale=0.48]{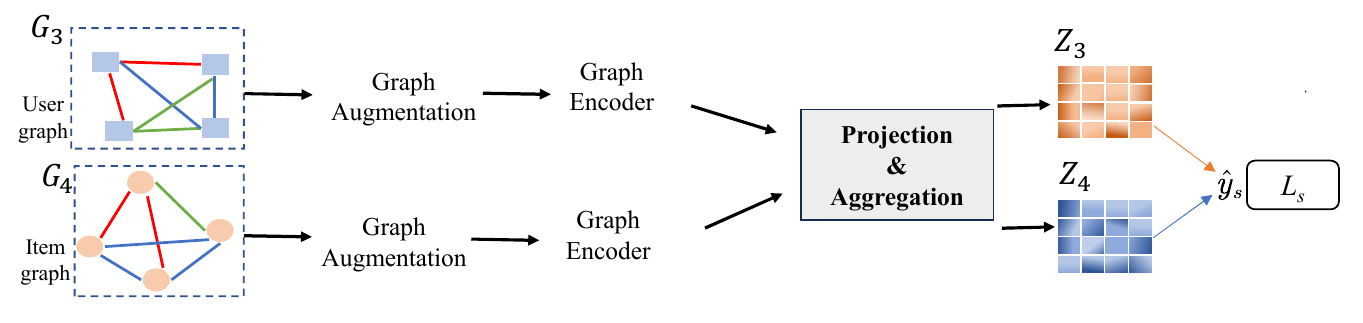}
    \caption{The subtask of MCGCL.}
    \label{fig:MCGCL2}
\end{figure*}

The subtask of MCGCL is illustrated in Fig. \ref{fig:MCGCL2}. Firstly, we identify hard samples by selecting the top $ \varepsilon $-fraction of samples with the highest entropy values. The entropy for each sample is computed as the cross-entropy loss from the main task, defined as:
$
\textit{Entropy}(G_n) = - y_{_{M_n}} \log \left( \hat{y}_{_{M_n}} \right) - (1 - y_{_{M_n}}) \log \left( 1 - \hat{y}_{_{M_n}} \right),
$
where $ y_{_{M_n}} $ and $ \hat{y}_{_{M_n}} $ are the true and predicted labels for the $ n $-th edge in the graph $ G $, respectively. The samples with the highest entropy values, corresponding to the hardest-to-predict edges, are selected as the hard samples for the new graph $ G_{\textit{hard}} $.
Formally, the hard sample graph $ G_{\textit{hard}} $ is defined as the set of samples with the top $ \varepsilon $-fraction of the highest entropy values:
$
G_{\textit{hard}} = \left\{ G_n \mid \textit{Entropy}(G_n) \in \mathcal{T}_{\varepsilon} \right\},
$
where $ \mathcal{T}_{\varepsilon} $ is the set of samples corresponding to the top $ \varepsilon $-fraction of the highest entropy values, defined as:
$
\mathcal{T}_{\varepsilon} = \left\{ G_n \mid \textit{Entropy}(G_n) \geq \mathcal{Q}_{1-\varepsilon} \left( \left\{ \textit{Entropy}(G_n) \mid G_n \in G \right\} \right) \right\}.
$
These hard samples, corresponding to the edges with the highest uncertainty, are then used for further processing and model refinement.

In the subtask, we particularly focus on the cold start problem. The cold start problem refers to the difficulty in accurately predicting due to the lack of historical data for this new task. In subtask prediction, our strategy is to transfer the node representations learned in the main task to the subtask to initialize the representations, thereby quickly establishing the initial model. Even if these nodes perform poorly in the main task and are identified as hard samples, we still use them as prior knowledge in the subtask to guide the task. 

Therefore, after identifying the hard samples, we analyze $ G_{\textit{hard}} $ in comparison to $ G $, creating the mask matrices $ M_1 $ and $ M_2 $. These matrices are generated as follows:
$
M_1 = \left\{ u \mid u \in \mathcal{U}_{G_{\textit{hard}}} \cap \mathcal{U}_G \right\},
$
$
M_2 = \left\{ i \mid i \in \mathcal{I}_{G_{\textit{hard}}} \cap \mathcal{I}_G \right\},
$
where $ \mathcal{U}_{G_{\textit{hard}}} $ and $ \mathcal{U}_G $ represent the sets of users in $ G_{\textit{hard}} $ and $ G $, respectively, and $ \mathcal{I}_{G_{\textit{hard}}} $ and $ \mathcal{I}_G $ represent the sets of items in $ G_{\textit{hard}} $ and $ G $, respectively. These matrices are used to select the corresponding representations from the main task.

These mask matrices are applied to the node representations learned in the main task, resulting in the extracted node representations in the subtask. Specifically, the mask operations are performed as follows:
\begin{equation}
    H_{\mathcal{_U}}^S = Z^M_{\mathcal{_U}} \odot M_1,
\end{equation}
\begin{equation}
H_{\mathcal{_I}}^S = Z^M_{\mathcal{_I}} \odot M_2,
\end{equation}
where $ H_{\mathcal{_U}}^S $ and $ H_{\mathcal{_I}}^S $ are the extracted representations for users and items in the subtask, respectively, and $ Z^M_{\mathcal{_U}} $ and $ Z^M_{\mathcal{_I}} $ are the node representations for users and items from the main task. The symbol $ \odot $ represents element-wise multiplication, which applies the mask matrices $ M_1 $ and $ M_2 $ to the node representations to select the corresponding elements. 
The resulting representations are then used to build the homogeneous subgraph for users and items, where:
 \begin{equation} \label{eq13}
        G_{\mathcal{_{U(I)}}}=\textit{softmax}(\textit{MLP}(H^S_{\mathcal{_{U(I)}}})\textit{MLP}( (H^{S}_{\mathcal{_{U(I)}}})^{T} ) 
\end{equation}
where $G_\mathcal{_U}$ and $G_\mathcal{_I}$ denote the homogeneous subgraphs for users and items. Following similar steps to the main task, graph augmentation techniques, such as edge addition or removal, are applied to generate augmented graphs $G_{\mathcal{_U}}^{'}$ and $G_{\mathcal{_I}}^{'}$. These augmented graphs are then classified by labels to extract representations via different encoders. We need to calculate the contrastive loss $L_{_{S_{p}}}$ for the augmented graph using the same encoder. This is followed by the projection using a two-layer \textit{MLP}, and the computation of the label-level contrastive loss $L_{_{S{_{c}}}}$ across different encoders. Subsequently, we perform the aggregation process, enabling us to derive the representations $Z^{S}_{\mathcal{_U}}$  and $Z^{S}_{\mathcal{_I}}$ that represent users and items. These representations are then utilized to execute link prediction tasks. 
The loss can be calculated as follows:
\begin{equation} \label{eq15}
    L_{_S} = \sum_{n=1}^{N^{\textit{hard}}} \left( y_{S_{n}} \log \left( \hat{y}_{S_{n}} \right) + \left\| Z^{M}_{\mathcal{_U}} \odot M_1 - Z^{S}_{\mathcal{_U}} \right\|_2^2 + \left\| Z^{M}_{\mathcal{_I}} \odot M_2 - Z^{S}_{\mathcal{_I}} \right\|_2^2 \right)
\end{equation}
In this formula, we apply cross-entropy loss to ensure predictive accuracy for the subtask. Additionally, a regularization term is included to reduce the discrepancy between the representations obtained in the subtask, $ Z^{S}_{\mathcal{_U}} $ and $ Z^{S}_{\mathcal{_I}} $, and those from the main task, $ Z^{M}_{\mathcal{_U}} $ and $ Z^{M}_{\mathcal{_I}} $.

\subsection{Multi-label link prediction}

In our framework, the main task is responsible for producing the representations of users and items, denoted as $Z^{M}_{\mathcal{_U}}$ and $Z^{M}_{\mathcal{_I}}$, while the subtask focuses on generating representations from hard samples, represented as $Z^{S}_{\mathcal{_U}}$ and $Z^{S}_{\mathcal{_I}}$.  Although the main task is more important, the features of the subtask are also invaluable, as they provide implicitly crucial information to aid in reasoning. To combine both sets of features, we introduce a weighted combination approach. Moreover, we employ the attention aggregation and getting the weight matrix,  $W^{S}_{\mathcal{_U}}$, $W^{M}_{\mathcal{_U}}$ and $W^{S}_{\mathcal{_I}}$, $W^{M}_{\mathcal{_I}}$ \cite{vaswani2017attention}. Then, we complete the combinations between the main task and subtask and obtain the final user and item representations, $Z_{\mathcal{_U}}$ and $Z_{\mathcal{_I}}$, as follows: 
\begin{equation} \label{eq17}
Z_{\mathcal{_U}}=Concat\left[ \left[ W^{S}_{\mathcal{_U}} \cdot Z^{S}_{\mathcal{_U}}+W^{M}_{\mathcal{_U}} \cdot(Z^{M}_{\mathcal{_U}}\odot M_1 )\right] ,Z^{M}_\mathcal{_U} \odot \left( \mathbf{I}-M_1 \right) \right] 
\end{equation}
\begin{equation} \label{eq18}
Z_{\mathcal{_I}}=Concat\left[ \left[W^{S}_{\mathcal{_I}} \cdot Z^{S}_{\mathcal{_I}}+W^{S}_{\mathcal{_I}} \cdot(Z^{M}_{\mathcal{_I}}\odot M_2 )\right] ,Z^{M}_{\mathcal{_I}}\odot \left( \mathbf{I}-M_2 \right) \right] 
\end{equation}
Here, $M_1$ and $M_2$ act as mask matrices that isolate the components common to both the main task and the subtask. The expressions $\mathbf{I} - M_1$ and $\mathbf{I} - M_2$, with $\mathbf{I}$ denoting the identity matrix, subtract the shared elements, yielding matrices that represent the parts specific to the main task. We use these representations to calculate the validation loss for the graph data as the same as $G_{\textit{train}}$. Furthermore, the loss $L_{v}$ is derived by applying cross-entropy calculations.  Finally, the total loss in our framework can be formulated as:
\begin{equation} \label{eq16}
L_t= (\alpha(L_{_{M_{p}}}+ L_{_{M_{c}}})+\beta L_{_{M}})+(\mu (L_{_{S_{p}}}+L_{_{S_{c}}})+\gamma  L_{_S})+ L_v
\end{equation}
where $\alpha$ and $\beta$ are hyperparameters for the main task and $\mu$ and $\gamma$ are hyperparameters for the subtask.
With the design of MCGCL, the training procedure can be summarized in Algorithm \ref{alg:algorithm3}. 

\begin{algorithm}
\small
\caption{MCGCL framework procedure}
\label{alg:algorithm3}
\begin{algorithmic}[1]
\renewcommand{\algorithmicrequire}{\textbf{Input:}}
\renewcommand{\algorithmicensure}{\textbf{Output:}}
\REQUIRE {bipartite graph $G$; epochs for main task $T_0$, subtask $T_1$, and validation $T_2$.
}
\ENSURE{final node representations $Z_{\mathcal{_U}}, Z_{\mathcal{_I}}$.
}
\
\STATE{initialize parameters for $GCN$ encoders dedicated to the validation task.}
\STATE{
execute the main task as outlined in Algorithm \ref{alg:main}, using the bipartite graph $G$ and main task epochs $T_{0}$.
}
\STATE{
retrieve the representations $Z^{M}_{\mathcal{_U}}$ and $Z^{M}_{\mathcal{_I}}$ produced by the main task.
}
\STATE{
construct user-user and item-item graphs, $G_{\mathcal{_{U(I)}}}$ and $G^{'}_{\mathcal{_{U(I)}}}$, utilizing representations from the main task.
}
\STATE{
perform the subtask using the homogeneous subgraphs $G_{\mathcal{_{U(I)}}}$ and $G^{'}_{\mathcal{_{U(I)}}}$ over $T_{1}$ epochs.
}
\STATE{
acquire the subtask representations $Z^S_{\mathcal{_U}}$ and $Z^S_{\mathcal{_I}}$ based on the homogeneous subgraph analysis.
}
\STATE{
integrate representations from the main task and subtask to derive the representations $Z_{\mathcal{_U}}$ and $Z_{\mathcal{_I}}$.
}
\RETURN { $Z_{\mathcal{_U}}$, $Z_{\mathcal{_I}}$}
\end{algorithmic}
\end{algorithm}

\subsection{Computational complexity}

We focus on the computational complexity in the sub-task. By selecting the hard sample parameter $ \varepsilon $, the computational complexity during the sub-task learning phase can be significantly reduced. For example, by setting $ \varepsilon$ = 0.3, only 30\% of the samples are processed as hard samples, which leads to a substantial decrease in computational requirements. Specifically, the complexity of contrastive loss and aggregation is reduced from $ O(|V|^2 d) $ to $ O((0.3 |V|)^2 d)$ = $O(0.09 |V|^2 d) $, resulting in a 90\% reduction. Similarly, the complexity of link prediction is reduced from $ O(N) $ to $ O(0.3 N_{\textit{hard}}) $, where $ N_{\textit{hard}} $ is the number of hard samples, leading to a 70\% reduction. This reduction in computational complexity demonstrates the effectiveness of using hard samples for sub-task processing. By focusing on the most challenging samples, we can reduce the overall computational burden, making the method more scalable and suitable for large-scale datasets.

\section{Experiments} \label{sec: experiments}

Our experiments were carried out on a system equipped with the NVIDIA GeForce RTX 4060 Ti GPU, with 16 GB of memory. Detailed information about the experimental setup is provided below. In this section, we assess MCGCL in link prediction across various real-world dataset sizes, aiming to address the following research questions:

\begin{itemize}
    \item \textbf{Q1 (Multi-label classification):} How effectively does MCGCL perform in the link prediction task within a multi-label context?
    
    \item \textbf{Q2 (Binary classification):} Is MCGCL capable of surpassing the performance of current leading methods, specifically SBGCL, in binary classification tasks?
    
    \item \textbf{Q3 (Ablation study):} How do the individual components of MCGCL contribute to its overall performance?
    
    \item \textbf{Q4 (Parameter analysis):} To what extent is MCGCL performance affected by its hyperparameters?
\end{itemize}

\subsection{Datasets}

Our experiments use six datasets from Amazon Reviews\footnote{http://snap.stanford.edu/data/web-Amazon-links.html}, namely Amazon, Arts, Automotive, Baby, Beauty, and Health. To reduce the impact of noisy data, we applied a filtering criterion to the nodes. Specifically, nodes with several connections falling below a frequency threshold of three were excluded. This could remove unreliable data points, ensuring dataset reliability for our analysis. Ratings were classified into three categories: ratings of 1 to 2 were \textit{Low}, 3 was \textit{Mid}, and 4 to 5 were \textit{High}. We used random sampling to divide the data. For each user, 80\% of the product data was randomly selected for training, 10\% for validation, and the remaining 10\% for prediction. This ensures that all users are included in the training, validation, and prediction phases. We conducted five crossover experiments on each dataset. Table \ref{table: datasets} presents the statistical details of the datasets used. In our study, we classified the Arts dataset as small-size, the Amazon dataset as large-size, and the other datasets as medium-size. For a fair comparison, we also selected the three most commonly used datasets in a signed bipartite graph \cite{zhang2023contrastive}, including Review, ML-1M, and Bonanza.

\begin{table}[ht] 
    \centering
     \caption{Statistical overview of the datasets. $|\textit{Edge}|$ denotes the total count of edges. $|\textit{User}|$ and $|\textit{Item}|$ indicates the number of user and item nodes. The \textit{Low}, \textit{Mid}, and \textit{High} reflect the distribution of edge labels.}
    \begin{tabular}{|c|c|c|c|c|c|c|}
        \hline
         \textbf{Dataset} & $|\textit{Edge}|$ & $|{\textit{User}}|$ & $|\textit{Item}|$ & \textit{Low} & \textit{Mid} & \textit{High} \\
         \hline
          Amazon & 668,334 & 312,929 & 22190 & 112,706 & 73509 & 482,119\\ 
         \hline
         Arts & 27750 & 24070 & 4207 & 4192 & 2209 & 21349\\ 
         \hline
         Automotive & 188,388 & 133,255 & 47540 & 28642 & 13635 & 146,111\\ 
         \hline
         Baby & 176,186 & 123,836 & 6941 & 30486 & 14783 & 130,917\\ 
         \hline
         Beauty & 248,872 & 167,724 & 28805 & 37595 & 17821 & 193,456\\ 
         \hline
         Health & 421,628 & 311,635 & 39276 & 69424 & 32074 & 320,130\\ 
         \hline
         Review & 1170 & 182 & 304 & 706 & --- & 464\\ 
         \hline
         ML-1M & 1,000,209 & 6040 & 3952 & 424,928 & --- & 575,281\\ 
         \hline
         Bonanza & 36,543 & 7919 & 1973 & 738 & --- & 35,805\\ 
         \hline
    \end{tabular}
    \label{table: datasets}
\end{table}

\subsection{Baselines and experiment setting}

We conduct various kinds of methods between our model, MCGCL, including, end-to-end-based, contrastive learning-based, and signed graph-based. End-to-end-based methods including M-GNN \cite{wang2019robust}, NeuLP \cite{zhong2020neulp} and SEAL \cite{cai2021line}. Contrastive learning based (excluding GNN) including C-GMAVE \cite{bai2022gaussian}, LRDG \cite{zhang2024multi}, MulSupCon \cite{zhang2024multi}. In our paper, we focus on signed GNN methods like SGCN \cite{derr2018signed}, SGCL \cite{shu2021sgcl}, SBGNN \cite{huang2021signed}, and SBGCL \cite{zhang2023contrastive}. We employed Pytorch and the torch\_geometric library for model execution. For optimization, we used the Adam optimizer with a learning rate of 0.005 and a weight decay of 1e-5. All methods, including MCGCL, were configured with a dimension size of 32, following in SBGCL \cite{zhang2023contrastive}. Graph augmentation involved random edge modifications with a perturbation probability of 1\%. The hyperparameters $\varepsilon$ for hard sampling were set to 0.3. Performance evaluation of MCGCL was conducted using four metrics in a multi-label classification task: AUC, Macro-F1, and Micro-F1. 

\subsection{Multi-label classification (Q1)}

The experimental results for the multi-label classification task are presented in Table \ref{tab:results}. MCGCL demonstrates outstanding performance across all datasets, showing significant improvements in various metrics. It consistently scores higher than the average of end-to-end methods and generally outperforms existing contrastive learning and Signed GNN methods. Specifically, the performance in AUC, Macro-F1, and Micro-F1 metrics is 10-17\% higher than end-to-end methods, 4-17\% higher than Signed GNN methods, and slightly higher than contrastive learning methods by approximately 1-3\%. Therefore, MCGCL exhibits strong capability in handling multi-label classification tasks, particularly in learning more precise node representations across different dataset sizes. For instance, on the Amazon dataset, the AUC is 12.16\% higher than the average of end-to-end methods, 16.87\% higher than Signed GNN methods, and 4.45\% higher than contrastive learning methods. Similar trends are observed across medium-size datasets. On the Arts dataset, MCGCL achieves an AUC of 77.67, which is 7.91\% higher than end-to-end methods, 9.32\% higher than Signed GNN methods, and 3.14\% higher than contrastive learning methods. These results highlight the robustness of learning precise node representations across datasets of varying sizes.

\begin{table*}[t]
\centering
\caption{Multi-label classification results (\%) on six datasets. \textbf{\small Bold} numbers denote the best results, and \underline{underline} numbers denote the top performers for each kind.}
\resizebox{\textwidth}{!}{
\begin{tabular}{lcccccccccccc}
\toprule
& \multicolumn{3}{c}{Amazon} & \multicolumn{3}{c}{Arts} & \multicolumn{3}{c}{Automotive}  \\ \cmidrule(lr){2-4} \cmidrule(lr){5-7} \cmidrule(lr){8-10}
 & \multicolumn{1}{c}{AUC} & \multicolumn{1}{c}{Macro-F1} & \multicolumn{1}{c}{Micro-F1} &
\multicolumn{1}{c}{AUC} & \multicolumn{1}{c}{Macro-F1} & \multicolumn{1}{c}{Micro-F1} &
\multicolumn{1}{c}{AUC} & 
\multicolumn{1}{c}{Macro-F1} & \multicolumn{1}{c}{Micro-F1}  \\
    \midrule
    M-GNN & 56.31 & 59.12 & 58.67 
    & 69.56 & 72.94 & 69.56 
    & 58.12 & 66.78 & 65.23  \\
  NeuLP & 57.39 & 60.23 & 59.56 
    & 72.89 & 73.25 & 72.67 
    & 59.23 & 67.89 & 66.34  \\
       SEAL & \underline{61.45} & \underline{64.34} & \underline{64.45} 
    & \underline{73.67} & \underline{76.28} & \underline{74.45} 
    & \underline{59.39} & \underline{68.30} & \underline{67.52}  \\
    \midrule
    C-GMAVE & 59.63 & 65.41 & 64.85 
    & 72.29 & 75.69 & 74.81 
    & 64.15 & 69.93 & 68.16 \\
    LRDG & 62.49 & \underline{67.80} & 68.39 
    & 74.75 & 79.39 & 75.26 
    & 63.81 & 68.15 & 67.58  \\
    MulSupCon & \underline{64.33} & 67.42 & \underline{68.69} 
    & \underline{75.83} & \underline{80.63} & \underline{80.93} 
    & \textbf{\small \underline{69.75}} & \textbf{\small \underline{73.74}} &  \textbf{\small \underline{73.21}}  \\
    \midrule
    SGCN & 54.78 & 64.62 & 63.81 
    & 66.44 & 78.85 & 78.20 
    & 62.58 & 66.33 & 65.85\\
    SGCL & 55.69 & 63.88 & 63.35 
    & \underline{75.33} & 77.31 & 78.41 
    & \underline{62.75} & \underline{68.27} & 67.25 \\
    SBGNN & \underline{57.78} & 65.38 & \underline{65.71} 
    & 70.35 & 75.85 & 76.35 
    & 57.33 & 65.89 & 65.13  \\
    SBGCL & 55.25 & \underline{65.55} & 65.15 
    & 72.65 & \underline{78.91} & \underline{78.55} 
    & 60.80 & 67.53 & \underline{67.85} \\
    \midrule
    MCGCL (triple) & \textbf{\small 65.48} & \textbf{\small 68.52} & \textbf{\small 69.17} & 
    \textbf{\small 77.67} & \textbf{\small 80.97} & \textbf{\small 81.23} 
    & 68.67 & 71.66 & 71.03  \\
    \bottomrule
    \toprule
    & \multicolumn{3}{c}{Baby} & \multicolumn{3}{c}{Beauty} & \multicolumn{3}{c}{Health} \\ \cmidrule(lr){2-4} \cmidrule(lr){5-7} \cmidrule(lr){8-10}
     & \multicolumn{1}{c}{AUC} & \multicolumn{1}{c}{Macro-F1} & \multicolumn{1}{c}{Micro-F1} &
    \multicolumn{1}{c}{AUC} & \multicolumn{1}{c}{Macro-F1} & \multicolumn{1}{c}{Micro-F1} &
    \multicolumn{1}{c}{AUC} &
    \multicolumn{1}{c}{Macro-F1} & \multicolumn{1}{c}{Micro-F1}  \\
    \midrule
     M-GNN & 58.38 & 63.17 & 64.62 
    & 64.70 & 73.14 & 72.56 
    & 60.25 & 66.79 & 65.27  \\
  NeuLP & \underline{56.63} & 60.23 & 59.40 
    & 64.84 & 73.25 & 72.61 
    & 61.38 & 67.83 & 66.35  \\
       SEAL & 57.58 & 61.37 & 60.49 
    & \underline{65.63} & 74.12 & 73.34 
    & \underline{62.44} & 68.62 & 67.40  \\
    \midrule
     C-GMAVE & 65.69 & 67.50 & 69.81 
    & 67.29 & 73.76 & 74.62 
    & 64.12 & 69.52 & 68.10 \\
    LRDG & \underline{69.48} & \underline{71.83} & \underline{73.56} 
    & \underline{71.73} & \underline{74.27} & \underline{75.39}
    & \textbf{\small \underline{70.66}} & \textbf{\small \underline{71.87}} & \textbf{\small \underline{72.03}}  \\
    MulSupCon & 68.36 & 70.96 & 71.75 
    & 70.83 & 73.59 & 72.61 
    & 63.73 & 68.59 & 70.66  \\
    \midrule
    SGCN & 51.25 & 57.84 & 58.18 
    & 52.68 & 61.33 & 61.46 
    & 52.33 & 65.51 & 64.05  \\
    SGCL & 54.38 & 62.61 & 62.31 
    & 55.74 & 63.18 & 62.98 
    & 52.83 & 63.86 & 64.32 \\
    SBGNN & 55.62 & \underline{63.57} & 62.18 
    & 56.65 & \underline{65.38} & 66.33 
    & \underline{62.35} &  \underline{69.53} & \underline{69.28}   \\
    SBGCL & \underline{60.67} & 63.53 & \underline{64.34} 
    & \underline{62.48} & 65.24 & \underline{66.43} 
    & 58.64 & 68.01 & 67.98 \\
    \midrule
    MCGCL (triple) & \textbf{\small 70.33} & \textbf{\small 73.93} & \textbf{\small 74.27} & 
    \textbf{\small 72.33} & \textbf{\small 75.25} & \textbf{\small 76.53} & 
    66.48 & 70.89 & 70.12   \\
    \bottomrule
\end{tabular}
}
\label{tab:results}
\end{table*}

\subsection{Binary classification (Q2)}

The evaluation results of the SBGCL and MCGCL for binary classification tasks are shown in Table \ref{tab:results1}. We conducted experiments on nine datasets. The result shows that MCGCL outperforms the SBGCL. The AUC value is larger than 0.7 on all datasets. Specifically, the AUC value for the Amazon and ML-1M datasets even shows an 8\% improvement compared to the SBGCL. SBGCL and MCGCL both demonstrated similar and excellent classification capabilities on small datasets such as Review, Arts, and Bonanza. Although SBGCL achieved a slightly better performance than MCGCL on the Review dataset, the improvement was not significant. However, for the larger datasets, MCGCL significantly outperforms SBGCL. This is likely because larger datasets tend to have more complex relationships, and MCGCL is better at handling such challenges.

\begin{table*}[t]
\centering
\caption{Binary classification performance (\%) on nine datasets. This table presents the comparison between MCGCL and SBGCL. \underline{Underlined} figures highlight the top-performing results.}
\resizebox{\textwidth}{!}{
\begin{tabular}{lcccc cccc cccc}
\toprule
& \multicolumn{3}{c}{Amazon} & \multicolumn{3}{c}{Arts} & \multicolumn{3}{c}{Automotive} \\ \cmidrule(lr){2-4} \cmidrule(lr){5-7} \cmidrule(lr){8-10} 
 & \multicolumn{1}{c}{AUC} & \multicolumn{1}{c}{Macro-F1} & \multicolumn{1}{c}{Micro-F1} &
\multicolumn{1}{c}{AUC} & \multicolumn{1}{c}{Macro-F1} & \multicolumn{1}{c}{Micro-F1} &
\multicolumn{1}{c}{AUC} & \multicolumn{1}{c}{Macro-F1} & \multicolumn{1}{c}{Micro-F1}  \\
\midrule
SBGCL 
& 62.52 & 66.04 & 65.36 
& 76.37 & 77.92 & 77.53 
& 72.59 & 73.70 & 73.14  \\
MCGCL (binary)
& \underline{70.84} & \underline{73.29} & \underline{72.47} 
& \underline{78.07} & \underline{79.82} & \underline{80.15} 
& \underline{75.90} & \underline{76.62} & \underline{77.31}  \\
\bottomrule
\toprule
& \multicolumn{3}{c}{Baby} & \multicolumn{3}{c}{Beauty} & \multicolumn{3}{c}{Health} \\ \cmidrule(lr){2-4} \cmidrule(lr){5-7} \cmidrule(lr){8-10} 
 & \multicolumn{1}{c}{AUC} & \multicolumn{1}{c}{Macro-F1} & \multicolumn{1}{c}{Micro-F1} &
\multicolumn{1}{c}{AUC} & \multicolumn{1}{c}{Macro-F1} & \multicolumn{1}{c}{Micro-F1} &
\multicolumn{1}{c}{AUC} & \multicolumn{1}{c}{Macro-F1} & \multicolumn{1}{c}{Micro-F1}  \\
\midrule
SBGCL 
& 69.94 & 72.84 & 72.44 
& 71.95 & 73.49 & 74.02 
& 65.97 & 68.05 & 68.87  \\
MCGCL (binary)
& \underline{74.62} & \underline{73.94} & \underline{69.09} 
& \underline{78.35} & \underline{78.71} & \underline{78.57} 
& \underline{72.75} & \underline{73.85} & \underline{72.14}  \\
\bottomrule
\toprule
& \multicolumn{3}{c}{Review} & \multicolumn{3}{c}{ML-1M} & \multicolumn{3}{c}{Bonanza} \\ \cmidrule(lr){2-4} \cmidrule(lr){5-7} \cmidrule(lr){8-10} 
 & \multicolumn{1}{c}{AUC} & \multicolumn{1}{c}{Macro-F1} & \multicolumn{1}{c}{Micro-F1} &
\multicolumn{1}{c}{AUC} & \multicolumn{1}{c}{Macro-F1} & \multicolumn{1}{c}{Micro-F1} &
\multicolumn{1}{c}{AUC} & \multicolumn{1}{c}{Macro-F1} & \multicolumn{1}{c}{Micro-F1}  \\
\midrule
SBGCL 
& 75.59 & 76.14 & 76.36 
& 63.67 & 66.04 & 65.73 
& 71.46 & 72.64 & 72.42  \\
MCGCL (binary)
& \underline{77.64} & \underline{79.11} & \underline{78.98} 
& \underline{71.25} & \underline{73.42} & \underline{73.35} 
& \underline{74.36} & \underline{76.74} & \underline{76.65}  \\
\bottomrule
\end{tabular}
}
\label{tab:results1}
\end{table*}

\subsection{Ablation study (Q3)} 

To assess the impact of various components in MCGCL, we conducted the ablation study, as shown in Table \ref{tb:abll}. Specifically, we evaluated the performance on small datasets Arts, medium datasets Beauty, and large datasets Amazon. We compared our method, MCGCL, with four variants: $MCGCL_{\textit{w/o}\ \textit{main task}}$, $MCGCL_{\textit{w/o}\ \textit{subtask}}$ and $MCGCL_{\textit{w/o}\ \textit{validation}}$. In $MCGCL_{\textit{w/o}\ \textit{main task}}$, it doesn't use the main task, and random representations are used to construct the homogeneous subgraph. The result shows that the AUC value is under 0.5 across these three datasets. This suggests that relying only on subtask learning may not be sufficient for MCGCL to classify instances with multi-label. When utilizing only the main task training on $MCGCL_{\textit{w/o}\ \textit{subtask}}$, we observe improvements across all datasets compared to $MCGCL_{\textit{w/o}\ \textit{main task}}$. For example, the AUC increases by 0.34 in the Arts, 0.33 in the Beauty, and 0.27 in the Amazon. This verifies the effectiveness of the main task. Furthermore, $MCGCL_{\textit{w/o validation}}$ represents the outcome of combining the main task and subtask without attention aggregation. The result continues to improve. Finally,  MCGCL represents the total results of our framework, and it achieves the best.

\begin{table}[t]
\scriptsize
\centering
\caption{Ablation studies of MCGCL.}
\label{tb:abll}
\setlength{\tabcolsep}{1.0em}
\begin{tabular}{lccc}
\toprule
 & \multicolumn{1}{c}{Arts} & \multicolumn{1}{c}{Beauty} & \multicolumn{1}{c}{Amazon} \\
 % & AUC & AUC & AUC \\
\midrule
w/o main task & 0.35 & 0.32 & 0.30 \\
w/o subtask   & 0.69 & 0.65 & 0.57 \\
w/o validation & 0.75 & 0.68 & 0.62 \\
MCGCL & 0.78 & 0.72 & 0.65 \\
\bottomrule
\end{tabular}
\end{table}

Data aggregation is frequently used in our framework. Particularly, aggregate representations from different augmented graphs and aggregate representations across different encoders. The choice of aggregation method can have a significant impact on our results. Two methods for aggregation in contrastive learning are \textit{MLP} and attention aggregation. To determine the most suitable method for our framework, we conducted experiments, as presented in Table \ref{tb:agg}. Firstly, we combine the representations with their mean value. This method produced the worst results. In addition, we utilized the \textit{MLP} to perform aggregation, which led to improved performance. However, the most effective approach was found to be the utilization of attention aggregation. Implementing attention led to further improvements in performance. Specifically, in the Arts, Beauty, and Amazon datasets, the AUC values improved by 0.05, 0.05, and 0.04.

\begin{table}[t]
\scriptsize
\centering
\caption{Ablation studies for aggregation with AUC.}
\label{tb:agg}
\setlength{\tabcolsep}{1.0em}
\begin{tabular}{lccc}
\toprule
 & \multicolumn{1}{c}{Arts} & \multicolumn{1}{c}{Beauty} & \multicolumn{1}{c}{Amazon} \\
 % & AUC & AUC & AUC \\
\midrule
Average combination & 0.65 & 0.60 & 0.56  \\
MLP & 0.73 & 0.67 & 0.61 \\
Attention & 0.78 & 0.72 & 0.65 \\
\bottomrule
\end{tabular}
\end{table}

\subsection{Hyper-parameter analysis (Q4)}

\begin{figure}[ht]
    \centering
    \includegraphics[clip,scale=0.50]{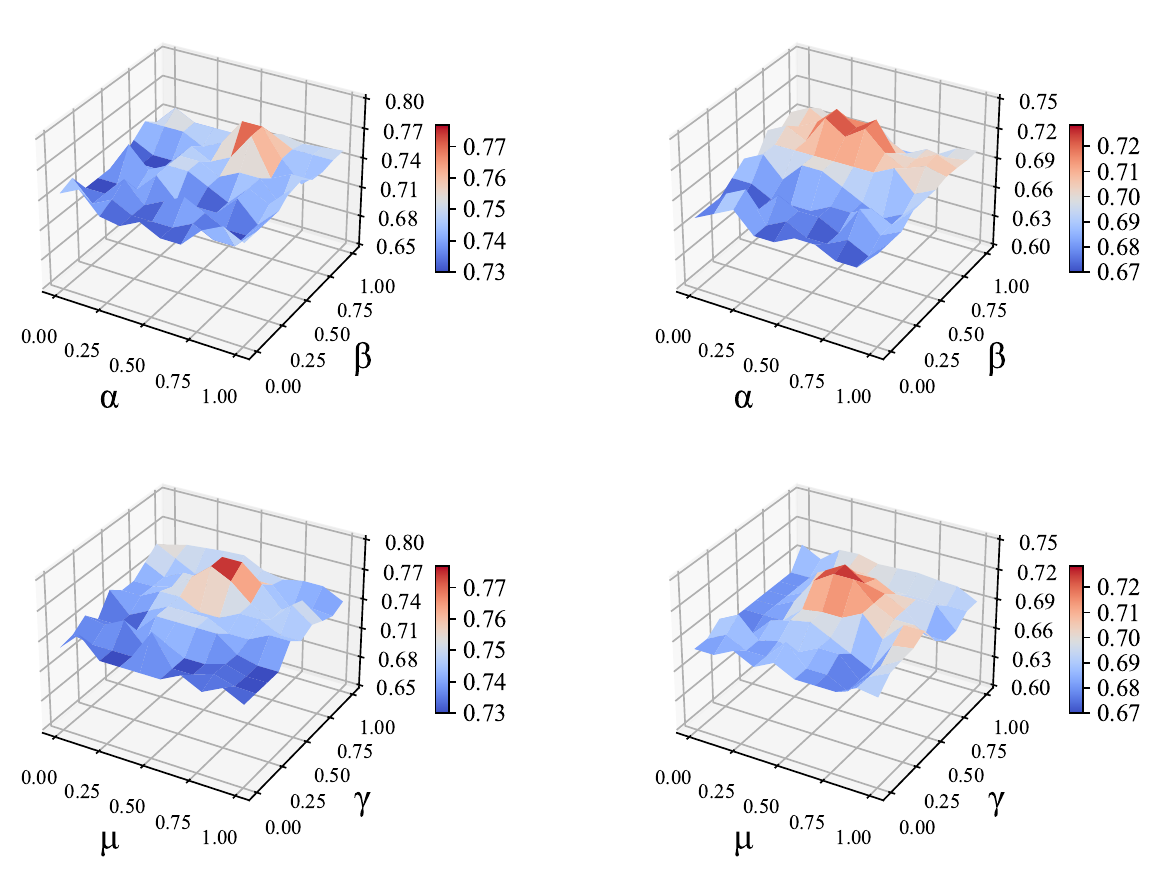}
    \caption{The AUC of Arts (left) and Beauty (right) with different parameters $\alpha$, $\beta$, $\mu$, and $\gamma$}
    \label{fig:3D}
\end{figure}

We conducted experiments on the Arts and Beauty datasets to analyze the impact of the $\alpha$, $\beta$, $\mu$, and $\gamma$ parameters in Equation \ref{eq16}. We explored different combinations of $\alpha$ and $\beta$, $\mu$ and $\gamma$, as shown in Fig. \ref{fig:3D}. For the Arts dataset, we observed that the AUC achieved its highest value of 0.78 when $\alpha$ was set to 0.6 or 0.7, and $\beta$ was set to 0.7 or 0.8. When $\alpha$ was around 0.9 and $\beta$ was approximately 0.2, it resulted in a lower AUC value. For the parameters $\mu$ and $\gamma$, the highest AUC values were reached when $\mu$ was 0.5 or 0.6 and $\gamma$ was 0.7 or 0.8. In contrast, when $\mu$ was 0.9 and $\gamma$ was 0.3, the AUC value was the lowest. This highlights the significance of the cross-entropy loss to the total loss. For the Beauty dataset, the highest value 0.72 was obtained when $\alpha$ was 0.4 or 0.6 and $\beta$ was 0.7 or 0.8. The best results were obtained when $\mu$ was 0.5 or 0.6 and $\gamma$ was 0.6 or 0.7. When $\alpha$ or $\mu$ was raised and $\beta$ or $\gamma$ was lowered, the AUC value decreased. Therefore, we chose the $\alpha$ and $\beta$ parameters to be 0.6 and 0.8 and the $\mu$ and $\gamma$ parameters to be 0.6 and 0.7, which enables MCGCL to reach the maximum AUC value.

\begin{figure}[ht]
    \centering
    \includegraphics[clip,scale=0.35]{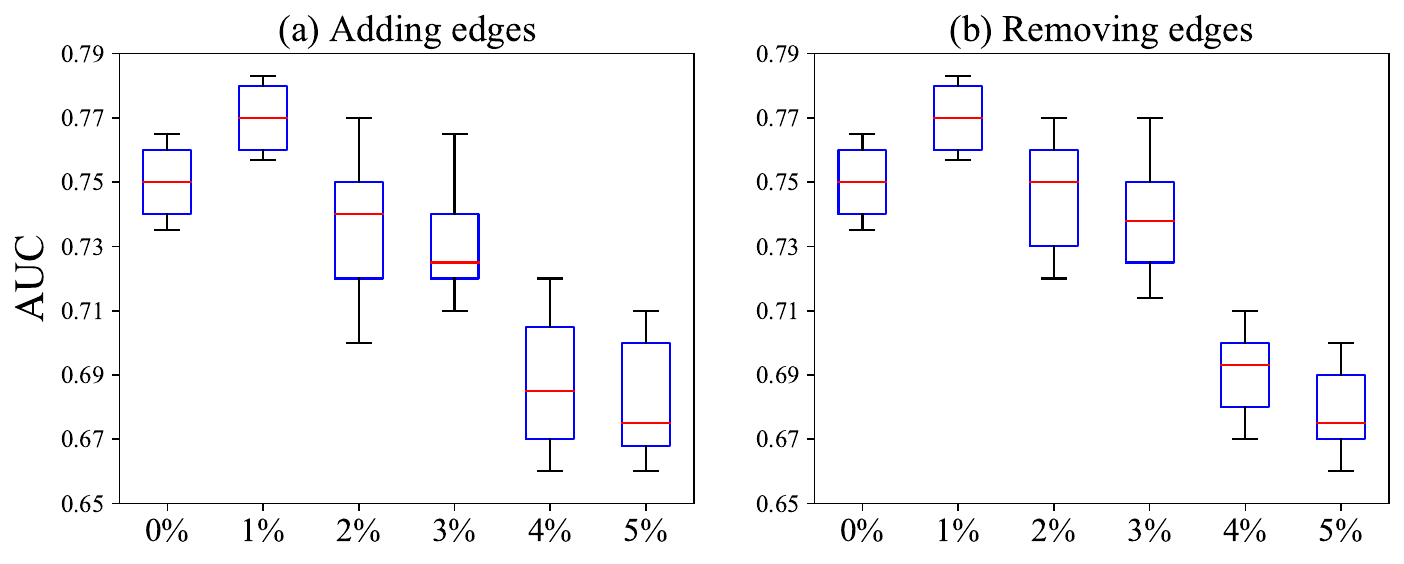}
    \caption{Effectiveness of the augmentation methods.}
    \label{fig:boxing}
\end{figure}

The most commonly used methods for graph augmentation in contrastive learning are edge adding and edge removing. To evaluate the effectiveness of these methods, we experiment on permutation probability range from 1\% to 5\% in the Arts dataset, as shown in Fig. \ref{fig:boxing}. Comparing with Fig. \ref{fig:boxing} (a) and (b), we observe that both adding and removing edges exhibit a similar changing trend. Moreover, we achieve the optimal performance with a medium AUC value of 0.77 when the probability is set to 1\%. When the probability is 0\%, the noise generated is minimal, resulting in the model being robust but not optimal. At 1\%, the noise level is moderate and beneficial for learning. The medium AUC value tends to decrease while the probability increases. When the probability is 1\%, the noise generated is relatively small, resulting in the model being more robust. As the probability increases, more noise is introduced, which leads to a lower AUC value.  In particular, when the probability is set to 5\%, there is a significant drop in the median AUC value, decreasing from 0.77 to 0.67, which is a gap of 0.1 compared to the setting of 1\%. Therefore, while a moderate amount of noise can enhance learning, excessive noise can disrupt the learning process. Meanwhile, we also found that adding edges introduces more noise, and removing edges produces better and more stable results. Therefore, the AUC value for edge removing is higher than for edge adding, but the difference is not significant.

\section{Conclusion} \label{sec: conclusion}

In the field of recommendation systems, accurately predicting user preferences and behaviors is crucial for delivering personalized experiences. The complexities in multi-label classification within bipartite graphs present unique challenges. Our proposed model, MCGCL, addresses these challenges through the innovative use of contrastive learning and multi-task strategies. We introduced MCGCL, a new approach designed to address the complexities of link prediction in bipartite graphs, particularly in multi-label contexts. By combining contrastive learning and multi-task strategies, MCGCL improves the accuracy of recommendations. Our framework uses a dual-phase learning process. It uses holistic bipartite graph learning for the main task and explores user-user and item-item views through homogeneous subgraph learning for the subtask. Extensive experiments on real-world datasets have demonstrated superior performance compared to existing state-of-the-art methods. This highlights its potential to enhance recommendation systems and its relevance in various business scenarios.

In the future, we plan to explore the integration of task division using teacher-student training \cite{wang2021knowledge}. We suspect that current attention aggregation may not optimally combine the two tasks. In the teacher-student framework, there is a mutual exchange of knowledge. Either participant can take on the role of the teacher, supporting the other. This interaction leads to better integration and learning outcomes. Additionally, we will explore to use more advanced training skills for large-scale graph data to improve learning efficiency and overall performance \cite{zheng2024structure}.

\section*{Acknowledgment}

This research was supported in part by the National Natural Science Foundation of China (No. 62272196), Guangzhou Basic and Applied Basic Research Foundation (No. 2024A04J9971), Natural Science Foundation of Guangdong Province (No. 2022A1515011861), Engineering Research Center of Trustworthy AI, Ministry of Education (Jinan University), and Guangdong Key Laboratory of Data Security and Privacy Preserving.

\bibliographystyle{ACM-Reference-Format}
\bibliography{MCGCL.bib}
\end{document}